%% file: IEEE-conference-template-062824.tex
\documentclass[conference, 11pt]{IEEEtran}
\IEEEoverridecommandlockouts

\usepackage{cite}
\usepackage{amsmath,amssymb,amsfonts}
\usepackage{algorithm}
\usepackage{algorithmic}
\usepackage{graphicx}
\usepackage{textcomp}
\usepackage{xcolor}
\usepackage{subfig}
\usepackage{titlesec}
\usepackage{bm,bbold}
\usepackage{caption}
\usepackage[T1]{fontenc}
\usepackage{newtxtext,newtxmath}
\DeclareMathOperator*{\argmin}{arg\,min}

\usepackage[letterpaper,
left=0.64in,
right=0.64in,
top=0.75in,
bottom=1.1in,
headheight=0.12in,
headsep=0.18in,
footskip=0.25in]{geometry}

\def\BibTeX{{\rm B\kern-.05em{\sc i\kern-.025em b}\kern-.08em
    T\kern-.1667em\lower.7ex\hbox{E}\kern-.125emX}}
\begin{document}

\title{Hierarchical Beam Training and Codebook Design for Movable Antenna-Assisted Near-Field Systems\\

}

\author{
\IEEEauthorblockN{
	Meihui Liu, 
	Qian Zhang, 
	Xuejun Cheng,
	Yuhui Jiao,
	Yunxiao Li,
	Ju Liu
} 
\IEEEauthorblockA{School of Information Science and Engineering, Shandong University, Qingdao, 266237, China \\
	Emails: \{meihuiliu, qianzhang2021, chengxuejun, yuhuijiao2024, yunxiaoli\}@mail.sdu.edu.cn, juliu@sdu.edu.cn}}
	\maketitle

\author{\IEEEauthorblockN{1\textsuperscript{st} Given Name Surname}
\IEEEauthorblockA{\textit{dept. name of organization (of Aff.)} \\
\textit{name of organization (of Aff.)}\\
City, Country \\
email address or ORCID}
\and
\IEEEauthorblockN{2\textsuperscript{nd} Given Name Surname}
\IEEEauthorblockA{\textit{dept. name of organization (of Aff.)} \\
\textit{name of organization (of Aff.)}\\
City, Country \\
email address or ORCID}
\and
\IEEEauthorblockN{3\textsuperscript{rd} Given Name Surname}
\IEEEauthorblockA{\textit{dept. name of organization (of Aff.)} \\
\textit{name of organization (of Aff.)}\\
City, Country \\
email address or ORCID}
\and
\IEEEauthorblockN{4\textsuperscript{th} Given Name Surname}
\IEEEauthorblockA{\textit{dept. name of organization (of Aff.)} \\
\textit{name of organization (of Aff.)}\\
City, Country \\
email address or ORCID}
\and
\IEEEauthorblockN{5\textsuperscript{th} Given Name Surname}
\IEEEauthorblockA{\textit{dept. name of organization (of Aff.)} \\
\textit{name of organization (of Aff.)}\\
City, Country \\
email address or ORCID}
\and
\IEEEauthorblockN{6\textsuperscript{th} Given Name Surname}
\IEEEauthorblockA{\textit{dept. name of organization (of Aff.)} \\
\textit{name of organization (of Aff.)}\\
City, Country \\
email address or ORCID}
}

\maketitle

\begin{abstract}
As sixth-generation (6G) communication systems evolve toward higher frequency bands and larger array apertures, the near-field range expands rapidly, making near-field channel estimation increasingly important and challenging. Beam training has been recognized as an effective approach for channel state information (CSI) acquisition. However, because of the propagation characteristics of spherical waves, beam training needs to perform a joint search in the angle and distance domains, which results in unaffordable beam training overhead. By flexibly reconfiguring antenna positions, movable antenna (MA) technology can fully exploit the spatial variations of wireless channels and achieve more accurate beam focusing, thereby providing additional flexibility for efficient beam training design. Therefore, based on MA-assisted near-field systems, we develop a hierarchical beam training strategy that combines reduced training overhead with high beam gain and design a corresponding hierarchical codebook. This codebook forms focused beams over the joint angle–distance domain, maximizing beam gain within the target region while suppressing energy leakage into non-target regions. Simulation results confirm substantial performance gains of the method over the conventional fixed-position antenna (FPA) system and the far-field beam training method.
\end{abstract}


\section{Introduction}
Multiple-input multiple-output (MIMO) technology enables efficient spectrum utilization together with fine spatial resolution, and serves as a key technology for improving the capacity and reliability of sixth-generation (6G) communication systems. However, under the conventional fixed-position antenna (FPA) architecture, the continuous expansion of antenna array size inevitably leads to considerable hardware expenditure, energy consumption, and signal-processing complexity, thereby constituting a bottleneck that constrains further improvements in wireless network performance. These limitations have motivated the use of movable antenna (MA) technology as an attractive solution. By permitting the antenna elements to reposition within a bounded spatial region, MA can utilize location-dependent channel variations across the continuous spatial domain, thereby achieving higher performance gains without requiring additional radio frequency (RF) chains~\cite{10286328,ding2024movable,10772590}.

When MA systems employ large spatial regions to exploit additional spatial reconfigurability, the enlarged effective array aperture increases the Rayleigh distance, making wireless propagation transition from the far-field regime toward the near-field regime~\cite{an2024near}. In this regime, electromagnetic waves exhibit spherical-wave characteristics instead of the plane-wave behavior typically assumed for far-field communications. As a consequence, conventional transmission schemes and channel models developed under the far-field plane-wave assumption are no longer suitable~\cite{10909572,yang2024near,9598863}. The considerable pilot and computational burden involved in estimating instantaneous channel state information (CSI) in large-scale antenna systems has driven the development of computationally lightweight and resource-efficient acquisition techniques~\cite{zhang2025multi}. Among these techniques, beam training is particularly attractive because it supports high signal-to-noise ratio (SNR) communication while requiring relatively limited training overhead, making it one of the most promising approaches for efficient CSI acquisition.

Unlike far-field beams, which primarily concentrate energy in a specific direction, near-field beams focus energy within a specific spatial region~\cite{9738442,10068140}. Consequently, traditional far-field beam training schemes that rely solely on angular information are no longer applicable in near-field environments. Accordingly, a variety of beam training techniques tailored to near-field propagation have been explored. Zhang \textit{et al.}~\cite{zhang2022fast} developed a two-stage procedure that reduces the search overhead by decomposing the two-dimensional search into separate stages for angle and distance estimation. More recently, joint angle-distance estimation has been achieved without constructing a dedicated near-field codebook by exploiting the beam-energy distribution of a conventional far-field discrete Fourier transform (DFT) codebook~\cite{wu2024near}. Despite these advances, existing hierarchical training methods have not provided a multi-resolution codebook that jointly covers the angle and distance domains. Furthermore, beam training and codebook design specifically tailored to MA-assisted near-field systems remain to be fully investigated.

In this paper, we exploit the spatial reconfigurability introduced by MA to further enhance near-field beamforming, achieve more accurate beam focusing, and address the problems of hierarchical beam training and multi-resolution codebook design in MA-assisted near-field systems. The main contributions are listed as follows.
\begin{itemize}
	\item We establish an MA-assisted near-field system model under spherical-wave propagation and develop a hierarchical beam training scheme along with a corresponding multi-resolution codebook covering both the angle and distance domains.                
	 
	\item An efficient block coordinate descent (BCD) algorithm is developed to alternately optimize the BS precoding vector and the antenna positions, thereby minimizing the beam pattern matching error. The BS precoding vector optimization subproblem is handled using the Lagrange multiplier method, while the highly nonconvex antenna position optimization subproblem is addressed via the projected gradient descent (PGD) method.
	
	\item The simulation results verify that the proposed codebook achieves better beam focusing performance than the FPA and far-field codebooks. Meanwhile, the hierarchical beam training scheme maintains performance close to exhaustive beam training with a considerably lower training overhead.
\end{itemize}
\begin{figure}[t]
	\centering
	{\includegraphics[width=7.8cm, height=7cm]{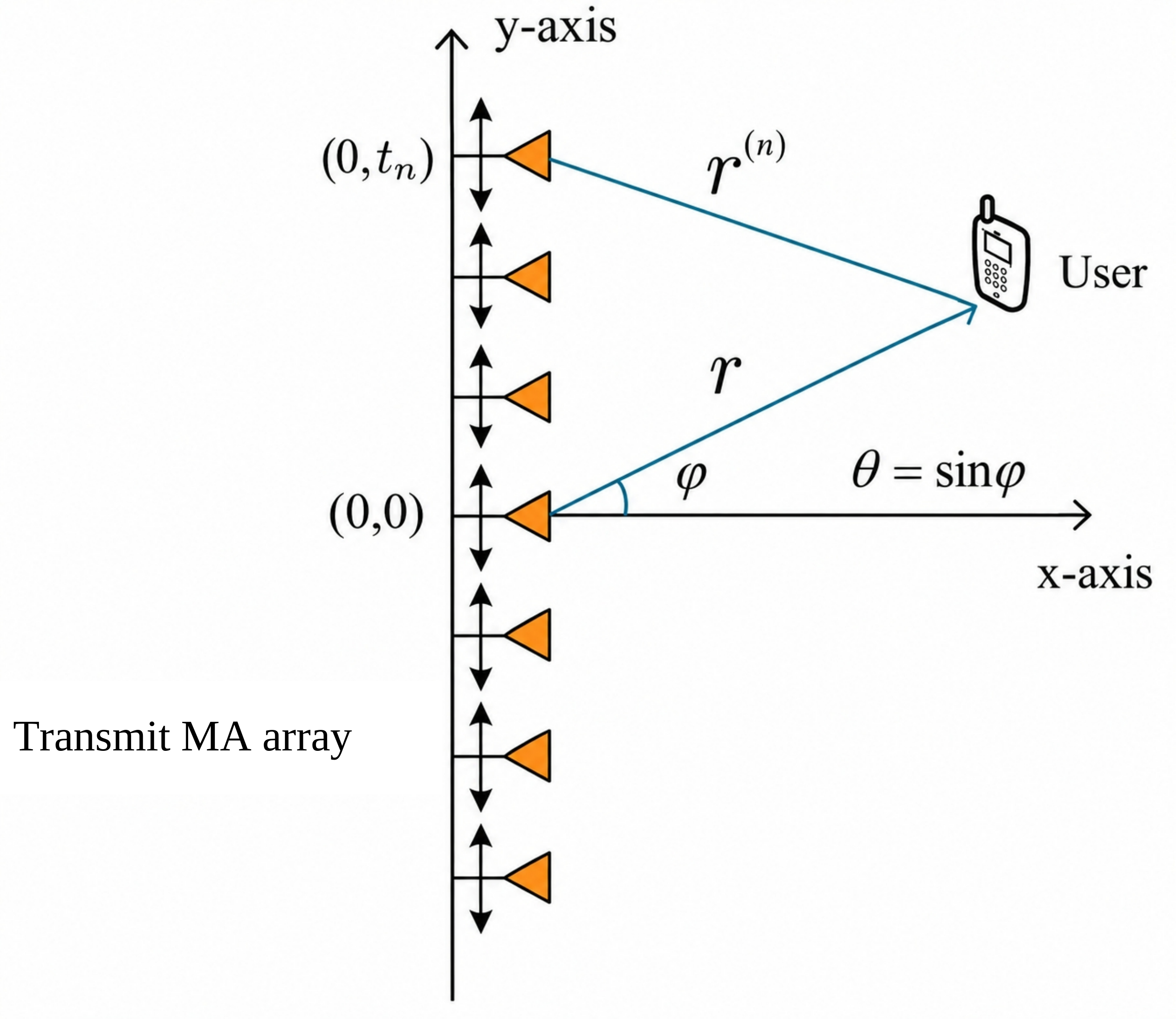}}
	\caption{The model of MA-assisted near-field communication systems.}
	\vspace{-0.4cm}
	\label{system_model}
\end{figure}
\section{System Model}
\subsection{Channel Model} 
The MA system considered for near-field beam training is illustrated in Fig. 1. A BS with $N$ MAs communicates with a single-antenna user. The MAs are allowed to translate over a one-dimensional aperture of length $L$. The $n$-th MA is located at $(0, t_n),  \text{where }t_n \in [-L/2, L/2]$. Using the uniform spherical wave (USW) propagation model~\cite{10220205}, the near-field line-of-sight (LOS) channel $\bm h$ between the BS and the user is written as
\begin{equation}
	\bm{h} = \sqrt{N}\alpha\bm{a}(\theta, r),
\end{equation}
where $\alpha = \frac{1}{\sqrt{4\pi r^2}}$ denotes the path loss from the center of the MA array to the user, and $r$ denotes the corresponding distance. The BS spatial angle is defined as $\theta = \sin \varphi$, where $\varphi$ denotes the angle of departure (AoD) from the center of the MA array towards the user. The near-field steering vector $\boldsymbol{a}(\theta, r)$ is expressed as
\begin{equation}
	\bm{a}(\theta, r) = \frac{1}{\sqrt{N}} \left[e^{-jk_c(r^{(1)} - r)}, \ldots, e^{-jk_c(r^{(N)} - r)} \right]^T,
\end{equation}
where $k_c = \frac{2\pi}{\lambda}$ denotes the wavenumber, and $\lambda$ denotes the corresponding wavelength. The propagation distance from the $n$-th MA to the user is denoted by $r^{(n)}$. Based on the Fresnel approximation~\cite{9723331}, $r^{(n)}$ can be expressed approximately as
\begin{equation}
	r^{(n)} \stackrel{}{\approx} r - t_n\theta + \frac{t_n^2 (1 - \theta^2)}{2r}.
\end{equation}

Consequently, the steering vector for the user can be reformulated as
\begin{equation}
	\bm{a}(\theta, r)\! \!=\!\! \frac{1}{\sqrt{N}}\!\begin{bmatrix}\!e^{-jk_c(\frac{t_1^2 (1 - \theta^2)}{2r}- t_1\theta)}\!,\! \ldots\!,\!e^{-jk_c\!(\frac{t_N^2 (1 - \theta^2)}{2r}- t_N\theta)} \end{bmatrix}^T.
\end{equation}

\subsection{Signal Model}
Let $s$ represent the BS transmit symbol. The resulting received signal at the user is then written as
\begin{equation}
	y = \bm{h}^H \bm{w}s + z = \sqrt{N}\alpha \bm{a}^H (\theta, r) \bm{w} s + z,
\end{equation}
where $\bm{w} \in \mathbb{C}^{N \times 1}$ is the precoding vector employed for BS transmission, while $z \sim \mathcal{CN}(0, \sigma^2)$ denotes the additive white Gaussian noise (AWGN) observed by the user.
\section{Near-Field Hierarchical Beam Training and Codebook Design}
\subsection{Hierarchical Beam Training}
Hierarchical beam training reduces training overhead by dividing the search into multiple levels. At the first level, a small number of wide beams cover the entire service region to obtain a coarse estimate of the user's location. At each subsequent level, narrower beams refine the search within the region selected at the preceding level until the user’s location is accurately identified. We assume that a total of $F$ levels are searched, and $R$ regions are searched in each level. Therefore, the hierarchical search requires $R \times F$ beam tests to determine the target user region. In contrast, if an exhaustive search is used, every region at the final level must be searched, corresponding to a training overhead of $R^F$. Therefore, we adopt a hierarchical beam training method with lower training overhead to obtain channel information and improve the effective transmission rate of the system.

\subsection{Hierarchical Codebook Design}
To accommodate beam training at different levels, we need to design codebooks with corresponding resolutions. The codeword design problem is formulated to optimize the actual beam pattern such that it approaches the desired ideal beam pattern as closely as possible, thereby obtaining the corresponding codeword.

Specifically, consider the codebook design at $l$-th level, where the coverage region is delineated by the angular interval $[\theta_{l, \min}, \theta_{l, \max}]$ and the distance interval $[r_{l, \min}, r_{l, \max}]$. Then, we perform sampling with $K$ points along the angular dimension and $S$ points along the distance dimension. The coordinates of the sampled grid points $(\theta_k, r_i)$ for the $l$-th level are determined by
\begin{align*}
	\theta_{k, l} &= \theta_{l, \min} + \left(k - \frac{1}{2}\right) \!\!\frac{\theta_{l, \max} - \theta_{l, \min}}{K}, \!\!\!\!\!\!\quad k = 1, 2, \dots, K, \\
	r_{i, l} &= r_{l, \min} + \left(i - \frac{1}{2}\right) \frac{r_{l, \max} - r_{l, \min}}{S}, \!\!\!\!\!\!\quad i = 1, 2, \dots, S.
\end{align*}

Let $P_{k,i}$ denote the desired amplitude gain at each sampling point $(\theta_k, r_i)$.  Mathematically, $P_{k,i}$ is given by
\begin{equation}
	P_{k,i} =
	\begin{cases}
		C_g, & \text{if } \theta_{k, l} \in \mathcal{D}_{\theta, l} \text{ and } r_{i, l} \in \mathcal{D}_{r, l}, \\
		0, & \text{otherwise},
	\end{cases}
\end{equation}
where $C_g > 0$ denotes the constant target gain desired within the target area, $\mathcal{D}_{\theta, l} \text{ and } \mathcal{D}_{r, l} $ denote the target angle and distance intervals. This ensures high gain within the target region while suppressing energy leakage into other areas.

Based on the sampling grid, the BS precoding vector $\bm{w}$ and the MA position vector $\bm{t}$ are jointly designed to match the desired beam pattern. At each sampling point $(\theta_k,r_i)$, the desired complex response is specified by the target amplitude $P_{k,i}$ and the phase $\phi_{k,i}$, where $\phi_{k,i}$ denotes an auxiliary phase-alignment term that is updated to match the phase of the actual beam response during the optimization process. Accordingly, the multi-resolution codebook optimization is expressed as
\begin{subequations}\label{eq:opt_prob}
	\begin{align}
		\min_{\bm{w}, \bm{t}} \quad & f(\bm{w}, \bm{t}) = \sum_{k=1}^K \sum_{i=1}^S \left| \bm{a}(\theta_k, r_i)^H \bm{w} - P_{k,i} e^{j\phi_{k,i}} \right|^2 \\
		\text{s.t.} \quad & \mathcal{C}_{\text{BS}}: \|\bm{w}\|_2^2 \leq P_{\max}, \label{eq:c_bs} \\
		& \mathcal{C}_{\text{FR}}: t_1 \geq -\frac{L}{2}, \quad t_N \leq \frac{L}{2}, \label{eq:c_fr} \\
		& \mathcal{C}_{\text{AC}}: t_n - t_{n-1} \geq D_0, \quad n = 2, 3, \dots, N. \label{eq:c_ac}
	\end{align}
\end{subequations}

In problem \eqref{eq:opt_prob}, constraint (7b) imposes the BS power budget, constraint (7c) confines the MA positions to $[-L/2,L/2]$, and constraint (7d) guarantees a minimum inter-antenna spacing of $D_0$.

\section{BS PRECODING AND ANTENNA POSITION OPTIMIZATION}
To handle the coupled variables in the objective function, we employ the BCD algorithm~\cite{10446199} to alternately update $\bm{w}$ and $\bm{t}$. Let $\ell$ denote the iteration index. The updates at the $(\ell+1)$-th iteration are formulated as
\begin{subequations}\label{eq:bcd_update}
	\begin{align}
		\bm{w}^{(\ell+1)} &= \arg \min_{\bm{w}} f\big(\bm{w}, \bm{t}^{(\ell)}\big) && \text{s.t.} \quad \mathcal{C}_{\text{BS}}, \label{eq:update_w} \\
		\bm{t}^{(\ell+1)} &= \arg \min_{\bm{t}} f\big(\bm{w}^{(\ell+1)}, \bm{t}\big) && \text{s.t.} \quad \mathcal{C}_{\text{FR}}, \mathcal{C}_{\text{AC}}. \label{eq:update_t}
	\end{align}
\end{subequations}
\subsection{BS Precoding Optimization}
For a given MA position vector, the BS precoding subproblem can be expressed as follows
\begin{equation}\label{eq:prob_w}
	\begin{split}
		\min_{\bm{w}}\quad & f(\bm{w}) = \sum_{k=1}^{K} \sum_{i=1}^{S} \left| \bm{a}(\theta_k, r_i)^H\bm{w} - P_{k,i}e^{j\phi_{k,i}} \right|^2 \\
		\text{s.t.} \quad & \|\bm{w}\|_2^2 \leq P_{\max}.
	\end{split}
\end{equation}

To facilitate the derivation, let $\mathbf{A} \in \mathbb{C}^{Q \times N}$ denote the steering matrix composed of the steering vectors $\bm{a}(\theta_k, r_i)$ with respect to all sampling points, and let $\bm{p} \in \mathbb{C}^{Q\times 1}$ denotes the vector containing the target response of these sampled locations, where $Q = K \times S$ represents the total number of samples.

Based on these definitions, the problem \eqref{eq:prob_w} admits the following compact matrix representation.
\begin{equation}\label{a-p}
	\min_{\bm{w}} \quad \|\mathbf{A}\bm{w} - \bm{p}\|_2^2 \quad \text{s.t.} \quad \|\bm{w}\|_2^2 \leq P_{\max}.
\end{equation}

The Lagrangian associated with problem (10) is 
\begin{equation}
	\mathcal{L}_{\bm{w}} = \|\mathbf{A}\bm{w} - \bm{p}\|_2^2 + \mu \left( \|\bm{w}\|_2^2 - P_{\max} \right),
\end{equation}
where $\mu \ge 0$ is the Lagrange multiplier corresponding to the power constraint.

The Karush-Kuhn-Tucker (KKT) conditions can be expressed as
\begin{subequations}
	\begin{align}
		& (\mathbf{A}^H\mathbf{A} + \mu \mathbf{I})\bm{w} - \mathbf{A}^H\mathbf{p} = 0; \\
		& \mu \ge 0; \|\bm{w}\|_2^2 \le P_{\max}; \mu \left( \|\bm{w}\|_2^2 - P_{\max} \right) = 0.
	\end{align}
\end{subequations}

According to the complementary slackness condition, the solution can be categorized into two cases: if $\bm{w}_u^H \bm{w}_u \leq P_{\max}$, then $\mu = 0$, and the optimal BS precoding is the unconstrained least-squares solution, $\bm{w}^\star = \bm{w}_u = (\mathbf{A}^H \mathbf{A})^{-1} \mathbf{A}^H \bm{p}$; If $\bm{w}_u^H \bm{w}_u > P_{\max}$, then $\mu > 0$, and the optimal solution must satisfy $\|\bm{w}^\star\|^2 = P_{\max}$, the optimal precoding vector is obtained in a closed-form manner as
	\begin{equation}\label{eq:w_opt_closed}
	\bm{w}^\star = (\mathbf{A}^H \mathbf{A} + \mu \mathbf{I})^{-1} \mathbf{A}^H \bm{p},
\end{equation}
where $\mu$ is obtained efficiently by a one-dimensional line search.

To avoid repeated matrix inversions when determining $\mu$, we exploit the eigenvalue decomposition (EVD) of the associated matrix, thereby reducing the computational complexity.

\subsection{Antenna Position Optimization}
Given the updated BS precoding vector $\bm{w}$, the original joint optimization problem is reduced to a subproblem with respect to the antenna position vector $\bm{t}$. Specifically, we minimize the mismatch between the synthesized beam response and the desired response by optimizing $\bm{t}$, subject to the constraints on the physical aperture $L$ and the minimum inter-element spacing $D_0$. The resulting subproblem is
\begin{equation}\label{eq:prob_t}
	\begin{aligned}
		\min_{\bm{t}} \quad & f(\bm{t}) = \sum_{k=1}^{K} \sum_{i=1}^{S} \left| \bm{a}(\theta_k, r_i)^H \bm{w} - P_{k,i}e^{j\phi_{k,i}} \right|^2 \\
		\textrm{s.t.} \quad & \bm{t} \in \mathcal{C}_{\text{FR}} \cap \mathcal{C}_{\text{AC}}.
	\end{aligned}
\end{equation}

Due to the non-linear phase terms associated with the antenna position $t_n$ in the steering vector $\bm{a}(\theta_k, r_i)$, $f(\bm{t})$ is highly non-convex. Consequently, we employ the PGD algorithm to optimize the antenna positions.

Define the residual term $e_{k,i}(\bm{t})\triangleq\bm{a}^H(\theta_k, r_i)\bm{w} - P_{k,i}e^{j\phi_{k,i}}$. An equivalent form of the objective function is $f(\bm{t}) = \sum_{k,i} e_{k,i}(\bm{t})e_{k,i}^*(\bm{t})$. Applying the chain rule, the derivative of $f(\bm{t})$ with respect to the $n$-th antenna position $t_n$ is given by
\begin{equation}\label{eq:partial_deriv}
	\frac{\partial f(\bm{t})}{\partial t_n} = 2 \sum_{k=1}^{K} \sum_{i=1}^{S} \text{Re}\left\{ e_{k,i}^*(\bm{t})\frac{\partial e_{k,i}(\bm{t})}{\partial t_n} \right\}.
\end{equation}

It is worth noting that within the residual term, only the steering vector $\bm{a}(\theta_k, r_i)$ exhibits dependency on $t_n$. Consequently, by differentiating the phase term of the steering vector, we derive the partial derivative with respect to $t_n$ as follows.
\begin{equation}\label{eq:steer_deriv}
	\frac{\partial [\bm{a}(\theta_k, r_i)]_n^*}{\partial t_n} = j k_c \left( \frac{1-\theta_k^2}{r_i}t_n - \theta_k \right) [\bm{a}(\theta_k, r_i)]_n^*.
\end{equation}

Substituting \eqref{eq:steer_deriv} into \eqref{eq:partial_deriv}, we finally obtain the analytical expression for the gradient of the objective function
\begin{equation}
	\begin{split}
		\frac{\partial f(\bm{t})}{\partial t_n} = & 2 \sum_{k=1}^{K} \sum_{i=1}^{S} \text{Re}\bigg\{  j k_c\cdot e_{k,i}^*(\bm{t}) \\
		& \times \left( \frac{1-\theta_k^2}{r_i}t_n - \theta_k \right) [\bm{a}]_n^* w_n \bigg\},
	\end{split}
\end{equation}
where $[\bm{a}]_n$ and $w_n$ denote the $n$-th elements of the steering vector and the precoding vector. Based on the partial derivative derived from each antenna, the full gradient vector of the antenna position vector $\bm{t} $ is constructed as
\begin{equation}
	\nabla_{\bm{t}} f(\bm{t}) = \left[ \frac{\partial f}{\partial t_1}, \dots, \frac{\partial f}{\partial t_N} \right]^T.
\end{equation}

Based on the derived gradient of the objective function, we update the MA positions along the negative gradient direction. Let $v$ denote the iteration index of the PGD algorithm.
\begin{equation}
	\tilde{\bm{t}}^{(v+1)} = \bm{t}^{(v)} - \eta^{(v)} \nabla_{\bm{t}} f(\bm{t}^{(v)}),
\end{equation}
where $\eta^{(v)}$ is selected using an Armijo backtracking line search so that each accepted step satisfies the prescribed descent condition.

To enforce the physical constraints on the updated antenna positions, the intermediate variable $\tilde{\bm{t}}^{(v+1)}$ is projected onto the feasible set $\mathcal{C}_\mathcal{\text{FR}} \cap \mathcal{C}_{\text{AC}}$.
\begin{equation}\label{eq:projection}
	\bm{t}^{(v+1)} = \Pi_{\mathcal{C}_{\text{FR}} \cap \mathcal{C}_{\text{AC}}}\left( \tilde{\bm{t}}^{(v+1)} \right) = \argmin_{\bm{t} \in \mathcal{\mathcal{C}_{\text{FR}} \cap \mathcal{C}_{\text{AC}}}} \|\bm{t} - \tilde{\bm{t}}^{(v+1)}\|_2^2.
\end{equation}
The projection in \eqref{eq:projection} constitutes a convex quadratic programming (QP) problem and can therefore be efficiently handled by conventional convex optimization tools, such as CVX.
		
		

\section{Simulation Results}
In the simulations, the BS employs $N=256$ MAs. The moving region of the antenna array is set to $L=300\lambda$, with the minimum inter-antenna spacing constrained to $D_0 = \lambda/2$. For the considered MA-assisted near-field system, the antenna movement region of length $L$ yields a Rayleigh distance of $\frac{2L^2}{\lambda}$. The noise power $\sigma^2$ and carrier frequency are configured as $-110$~dBm and $30$~GHz. The angle and distance sampling intervals are set to $[-1, 1]$ and $(20\text{m}, 50\text{m})$, with user locations distributed within the considered sampling region. The required beam gain $C_g$ is determined based on system requirements. We set it to 1 in this paper. Furthermore, the hierarchical beam training framework consists of $3$ levels, comprising $4$, $16$, and $64$ sub-regions, leading to a total of 12 searches. In contrast, the exhaustive beam training scheme needs to search all regions in the third level, thereby requiring a total of 64 searches. In the simulation experiments, under the single-user setting, the CSI associated with the codeword yielding the maximum achievable rate $R = \log_2 \left( 1 + \frac{|\bm{h}^H\bm{w}|^2}{\sigma^2} \right)$ is taken as the estimated CSI of the user.
\begin{figure}[h]
	\centering
	{\includegraphics[width=0.68\linewidth, height=5cm]{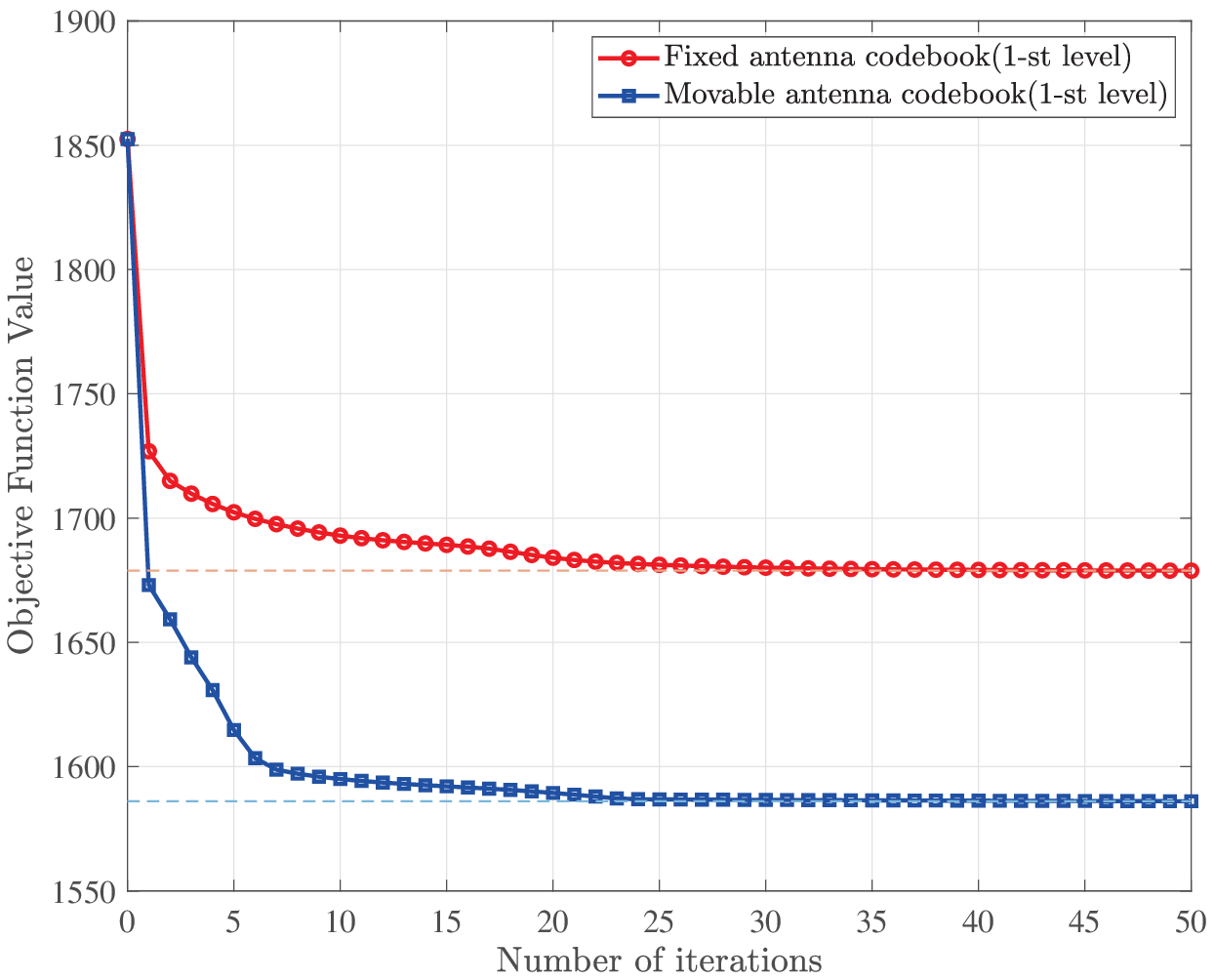}}
	\caption{Convergence performance of the proposed algorithm.}
	\vspace{-0.4cm}
	\label{Convergence behavior}
\end{figure}

Fig.~\ref{Convergence behavior} shows the objective values of the MA and FPA codebooks versus the iteration number for a first-level subregion. Both curves decrease monotonically and converge within a few iterations, indicating the fast convergence of the proposed optimization procedure. Moreover, the MA codebook converges to a significantly lower objective value than the FPA codebook, highlighting the performance benefit brought by the additional spatial degrees of freedom (DoFs) afforded by antenna mobility. By reconfiguring the antenna positions within the allowable movement region, the MA system improves near-field beam focusing.
\begin{figure}[t]
	\centering
	{\includegraphics[width=0.68\linewidth, height=5cm]{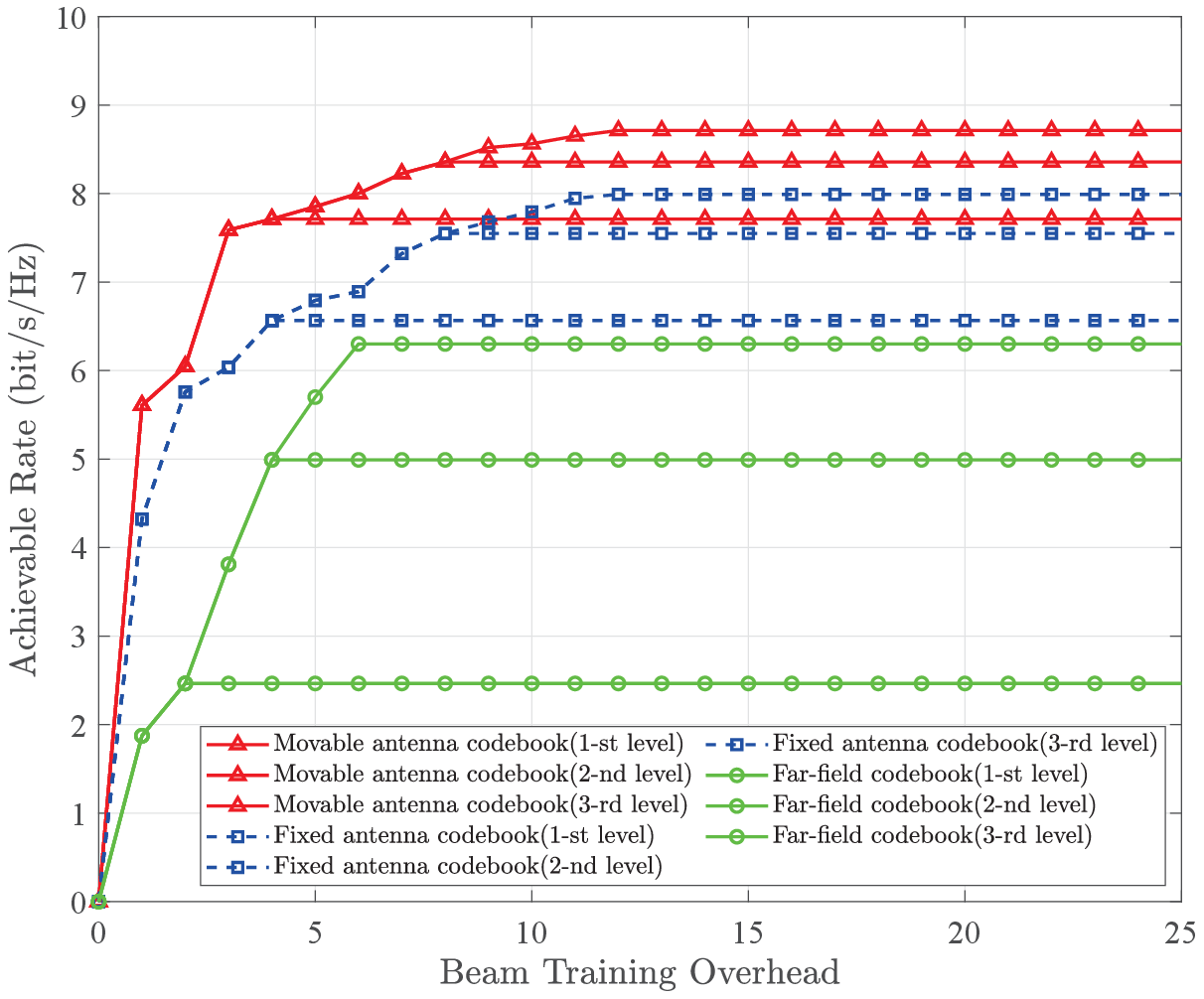}}
	\caption{Performance of three-level beam training with the MA codebook, FPA codebook and far-field codebook.}
	\vspace{-0.4cm}
	\label{MA codebook, FPA codebook and far-field codebook}
\end{figure}

\begin{figure}
	\centering
	{\includegraphics[width=0.68\linewidth, height=5cm]{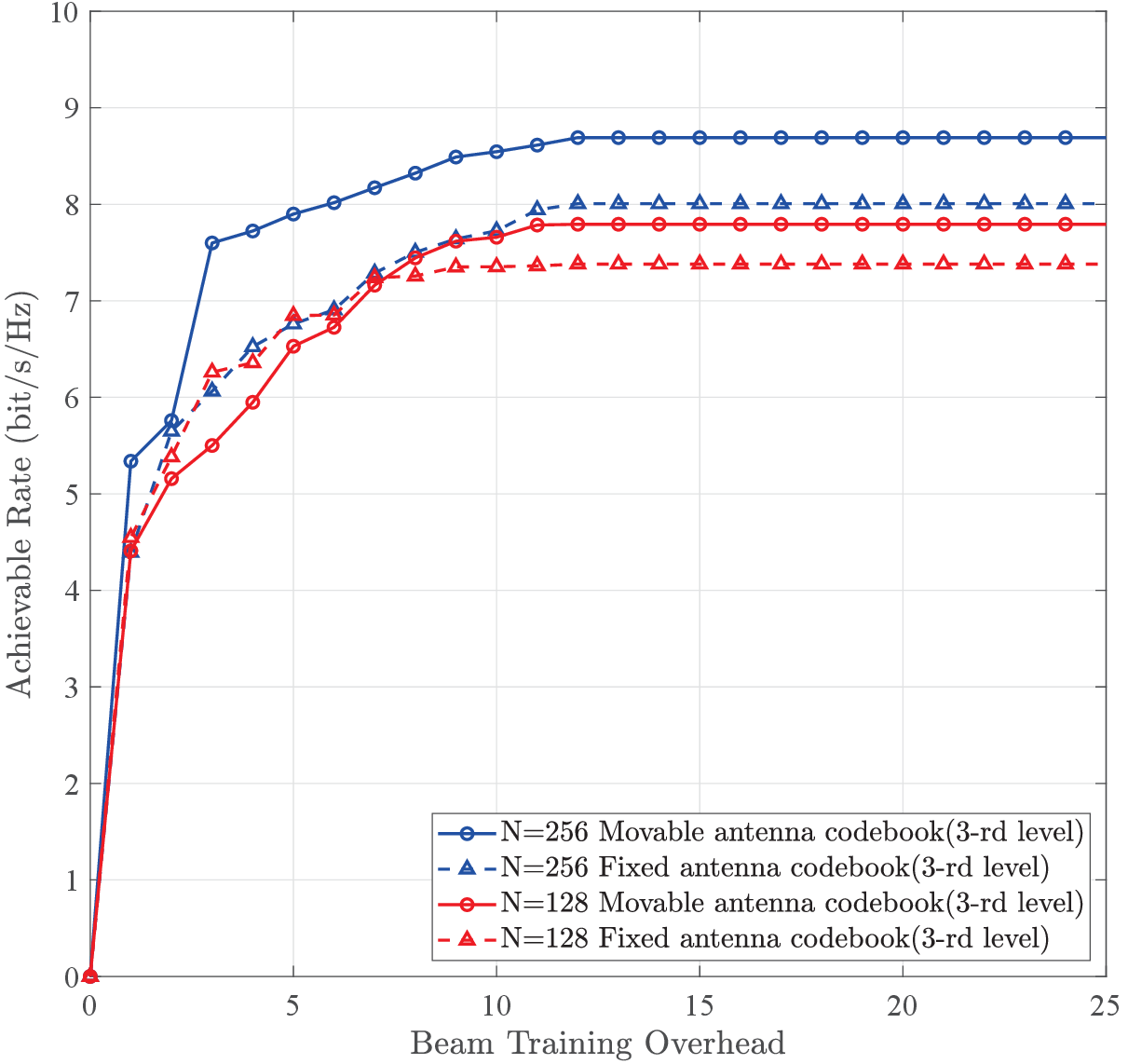}}
	\caption{Performance comparison under different schemes for the third-level beam training process.}
	\vspace{-0.4cm}
	\label{$M=256$ and $M=128$}
\end{figure}

Fig.~\ref{MA codebook, FPA codebook and far-field codebook} presents an achievable rate comparison among the MA, FPA, and far-field codebooks under three-level hierarchical beam training. The rate increases with the cumulative number of tested codewords. The MA codebook achieves the best near-field performance, whereas the far-field codebook performs the worst because its plane-wave assumption causes significant phase mismatch. In addition, the performance can be further improved by beam training in the subsequent stages within the range determined in the previous stage. These observations confirm that the proposed multi-resolution codebook design supports accurate user localization.

\begin{figure}
	\centering
	{\includegraphics[width=0.68\linewidth, height=5cm]{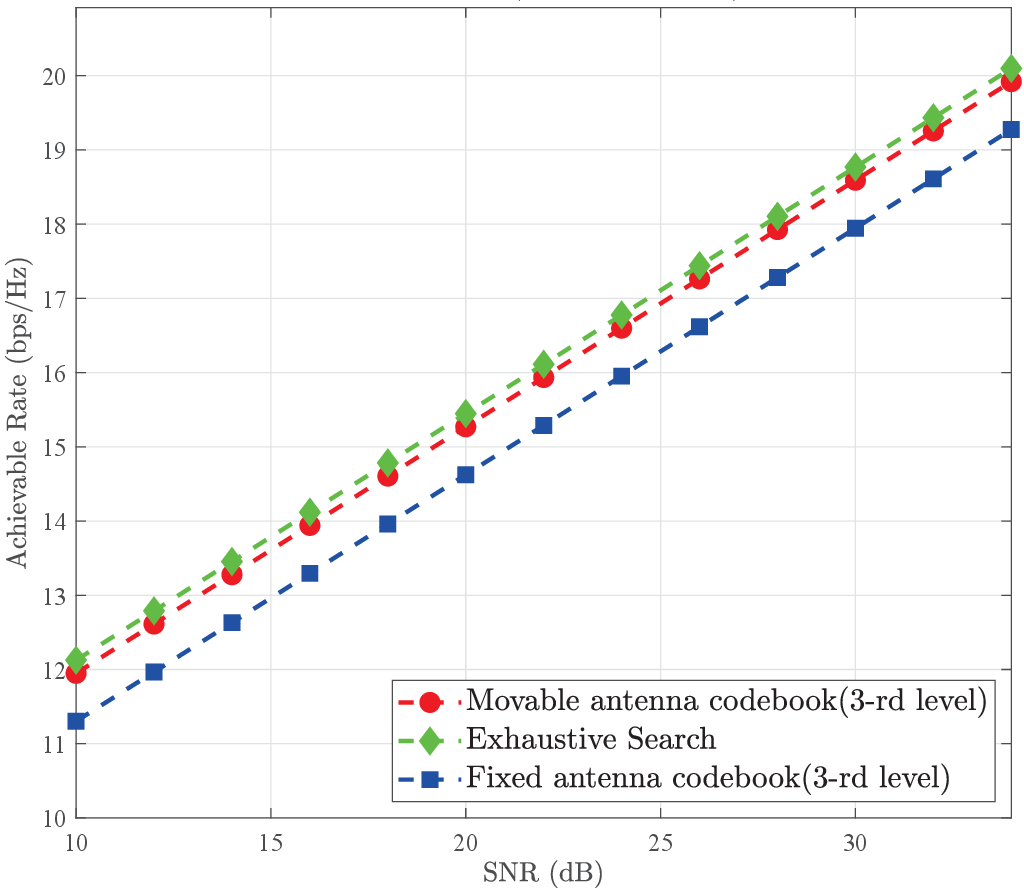}}
	\caption{Comparison of achievable rate performance versus SNR for the MA codebook, FPA codebook and exhaustive search.}
	\vspace{-0.4cm}
	\label{MA codebook, FPA codebook and exhaustive search}
\end{figure}

Fig.~\ref{$M=256$ and $M=128$} compares the achievable rate versus beam training overhead for different numbers of BS antennas. For the same training overhead, increasing \(N\) from 128 to 256 improves the achievable rate owing to the increased array gain and finer near-field focusing capability. Furthermore, the achievable rate obtained with the MA codebook is higher than that of the FPA codebook under the same number of antennas.

Fig.~\ref{MA codebook, FPA codebook and exhaustive search} compares the proposed MA codebook, the FPA codebook, and exhaustive search over different SNR levels. The proposed MA codebook closely approaches the exhaustive-search upper bound while requiring substantially less training overhead, demonstrating the effectiveness and efficiency of the hierarchical codebook design. 

\section{Conclusion}

In this paper, we considered MA-enabled near-field transmission and developed a hierarchical beam training scheme with low training overhead. Specifically, a multi-resolution codebook was designed to enable accurate beam focusing over the joint angle-distance domain. Simulation results showed that the proposed MA codebook provides substantial performance gains over both conventional FPA and far-field codebooks. These gains mainly arise from the additional spatial degrees of freedom introduced by MAs, enabling flexible antenna-position reconfiguration for improved system performance. Furthermore, the results confirm that the proposed hierarchical strategy achieves an achievable rate closely approaching the upper bound attained by exhaustive search, while requiring substantially less training overhead, thereby effectively balancing system performance and implementation complexity.

\bibliographystyle{IEEEtran}
\bibliography{refs}

\end{document}